\documentclass[twocolumn,twoside,slac_two]{revtex4}
\usepackage{graphicx}
\usepackage{fancyhdr}
\renewcommand{\headrulewidth}{0pt}
\renewcommand{\footrulewidth}{0pt}

\newcommand{\be}[0]{\begin{equation}}
\newcommand{\ee}[0]{\end{equation}}
\newcommand{\ba}[0]{\begin{eqnarray}}
\newcommand{\ea}[0]{\end{eqnarray}}

\def\ie{{\it i.e.}}

\begin{document}

\title{An approach to NNLO QCD analysis of non-singlet structure function}

%

\author{Ali N. Khorramian}
\affiliation{Physics Department, Semnan University, Semnan, Iran;
\\
Institute for Studies in Theoretical Physics and Mathematics
(IPM), P.O.Box 19395-5531, Tehran, Iran\\}

\author{S. Atashbar Tehrani}
\affiliation{Physics Department, Persian Gulf University, Boushehr,
Iran;
\\
Institute for Studies in Theoretical Physics and Mathematics (IPM),
P.O.Box 19395-5531, Tehran, Iran\\}

\begin{abstract}
We use  the next-to-next-to-leading order (NNLO) contributions to
anomalous dimension governing the evolution of non-singlet quark
distributions. We use  the $xF_3$ data of the CCFR collaboration to
obtain some unknown parameters which exist in the non-singlet quark
distributions in the NNLO approximation. In the fitting procedure,
Bernstein polynomial method is used. The results of valence quark
distributions in the NNLO, are in good agreement with the available
theoretical model.

\end{abstract}

\maketitle

\thispagestyle{fancy}


\section{Introduction}\label{section1}
The global parton analyses of deep inelastic scattering (DIS) and
the related hard scattering data are generally performed at NLO
order. Presently the next-to leading order (NLO) is the standard
approximation for most important processes in QCD.

The corresponding one- and two-loop splitting functions have been
known for a long time
\cite{Gross:1973rr,Georgi:1974sr,Altarelli:1977zs,Floratos:1977au,%
Floratos:1979ny,Gonzalez-Arroyo:1979df,Gonzalez-Arroyo:1980he,%
Curci:1980uw,Furmanski:1980cm,Floratos:1981hs,Hamberg:1992qt}. The
NNLO corrections should to be included in order to arrive at
quantitatively reliable predictions for hard processes occurring at
present and future high-energy colliders. These corrections are so
far known only for the structure functions in the deep-inelastic
scattering
\cite{ vanNeerven:1991nn, Zijlstra:1991qc,Zijlstra:1992kj,%
Zijlstra:1992qd}.\\

Recently much effort has been invested in computing NNLO QCD
corrections to a wide variety of partonic processes
 and therefore it is needed to generate parton
distributions also at NNLO so that the theory can be applied in a
consistent manner.\\

S. Moch and {\it{et al}}. \cite{Moch:2004pa,Vogt:2004ns}  computed
the higher order contributions up to three-loops splitting functions
governing the evolution of unpolarized non-singlet quark densities
in perturbative QCD.

During the recent years the interest to use CCFR data
\cite{CCFR:1997} for $xF_3$ structure function in higher orders,
based on orthogonal polynomial expansion method has been increased
\cite{Kataev:1997nc,
Kataev:1998ce,Alekhin:1998df,Kataev:1999bp,Kataev:2001kk,Santiago:2001mh}.

 In this paper we determine the flavor non-singlet
parton distribution functions, $xu_v(x,Q^2)$ and $xd_v(x,Q^2)$,
using the Bernstein polynomial approach up to NNLO level. This
calculation is possible now, as the non-singlet anomalous dimensions
in $n$-Moment space in three loops have been already introduced
\cite{Moch:2004pa,Vogt:2004ns}.

\section{The theoretical background of the QCD analysis}
In the NNLO approximation the deep inelastic coefficient functions
are known and also the anomalous dimensions in $n$-Moment space are
available at this order\cite{Moch:2004pa,Vogt:2004ns}. Since in this
paper we want to calculate the non-singlet parton distribution
functions in the NNLO by using CCFR experimental data, we introduce
the moments of non-singlet structure function up to three loops
order.

Let us define the Mellin moments for the $\nu$N structure function
$xF_3(x,Q^2)$:
\begin{equation}
\label{eq:momxf3IV} {\cal{M}}_n(Q^2)=\int_0^1
x^{n-1}F_3(x,Q^2)dx\;.
\end{equation}

The theoretical expression for these moments obey the following
renormalization group equation
\begin{equation}
\label{eq:RGE}
\bigg(\mu\frac{\partial}{\partial\mu}+\beta(A_s)\frac{\partial}
{\partial A_s} +\gamma_{NS}^{(n)}(A_s)\bigg)
{\cal{M}}_n(Q^2/\mu^2,A_s(\mu^2))=0\;, \label{rg}
\end{equation}
where $A_s=\alpha_s/(4\pi)$ is the renormalization group coupling
and is governed by the QCD $\beta$-function as
\begin{eqnarray}
\label{eq:BetaFun} \mu\frac{\partial
A_s}{\partial\mu}=\beta(A_s)=-2\sum_{i\geq 0} \beta_i A_s^{i+2}\;.
\end{eqnarray}
The solution of Eq.(\ref{eq:BetaFun}) in the NNLO is given by
\begin{eqnarray}
\label{eq:alfaNNLO}
A_s^{NNLO} &=&\frac{1}{\beta _{0}\ln Q^{2}/\Lambda _{\overline{MS}%
}^{2}}-\frac{\beta _{1}\ln (\ln Q^{2}/\Lambda
_{\overline{MS}}^{2})}{\beta
_{0}^{3}(\ln Q^{2}/\Lambda _{\overline{MS}}^{2})^{2}}+  \nonumber \\
&&\frac{1}{\beta _{0}^{5}(\ln Q^{2}/\Lambda
_{\overline{MS}}^{2})^{3}}[\beta _{1}^{2}\ln ^{2}(\ln
Q^{2}/\Lambda _{\overline{MS}}^{2})-\nonumber \\
&& \beta _{1}^{2}\ln (\ln Q^{2}/\Lambda
_{\overline{MS}}^{2})+\beta _{2}\beta _{0}-\beta _{1}^{2}]\;.\nonumber \\
\end{eqnarray}

Notice that in the above the numerical expressions for $\beta_0$,
$\beta_1$ and $\beta_2$ are
\begin{eqnarray}
\label{eq:beta}
\beta_0&=&11-0.6667f \;, \nonumber \\
\beta_1&=&102-12.6667f \;,\nonumber \\
\beta_2&=&1428.50-279.611f+6.01852f^2\;,
\end{eqnarray}
where $f$ denotes the number of active flavors.\\

 The solution of the renormalization group equation for non-singlet structure function $xF_3$ can be
presented in the following form \cite{Kataev:1999bp}:

\begin{equation}
\label{eq:momns} \frac{{\cal{M}}_n(Q^2)}{{\cal{M}}_n(Q_0^2)}=
exp\bigg[-\int_{A_s(Q_0^2)}^{A_s(Q^2)} \frac
{\gamma_{NS}^{(n)}(x)}{\beta(x)}dx\bigg]
\frac{C_{NS}^{(n)}(A_s(Q^2))} {C_{NS}^{(n)}(A_s(Q_0^2))}\;,
\end{equation}
where ${\cal{M}}_n(Q_0^2)$ is a phenomenological quantity related to
the factorization scale dependent factor which we will parameterize
in next section. $\gamma_{n}^{NS}$ is the anomalous function and has
a perturbative expansion as
 \be
\label{eq:gamma}
 \gamma_{n}^{NS}(A_s) =\sum_{i\geq
0} \gamma_{i}^{NS}(n) A_s^{i+1}\;. \ee At the NNLO the expression
for the coefficient function $C^{(n)}_{NS}$ can be presented as
\cite{Moch:1999eb}
\begin{equation}
\label{eq:coef}
C_{NS}^{(n)}(A_s)=1+C^{(1)}(n)A_s+C^{(2)}(n)A_s^2\;.
\end{equation}
With the corresponding expansion of the anomalous dimensions, given
by Eq.~(\ref{eq:gamma}), the solution to the three loops evolution
equation from Eq.~(\ref{eq:momns}), is as follows
\begin{eqnarray}
\label{eq:mnsNNLO} &&{\cal{M}}_n^{NNLO}(Q^2)=\left(
\frac{A_s(Q^{2})}{A_s(Q_{0}^{2})}\right) ^{\gamma _{0}^{NS}/2\beta
_{0}} \times \nonumber \\
&&\left\{1+ [A_s(Q^{2})-A_s(Q_{0}^{2})]\left(\frac{\gamma
_{1}^{}}{2\beta _{1}}-\frac{\gamma _{0}^{NS}}{2\beta _{0}}\right) \frac{%
\beta _{1}}{\beta _{0}} \right. \nonumber \\
&&+[A_s(Q^{2})-A_s(Q_{0}^{2})]^{2}\frac{\beta _{1}^{2}}{8\beta
_{0}^{2}}\left(
\frac{\gamma _{1}^{}}{\beta _{1}}-\frac{\gamma _{0}^{NS}}{\beta _{0}}%
\right) ^{2} \nonumber \\
&& +\frac{1}{4}[A_s^{2}(Q^{2})-A_s^{2}(Q_{0}^{2})]
\nonumber \\
&&
\left. \left( \frac{1}{\beta _{0}}%
\gamma _{2}^{}-\frac{\beta _{1}}{\beta _{0}^{2}}\gamma _{1}^{}+%
\frac{\beta _{1}^{2}-\beta _{2}\beta _{0}}{\beta _{0}^{3}}\gamma
_{0}^{NS}\right) \right \}\nonumber \\
&&\left(1+C^{(1)}A_s(Q^2)+C^{(2)}A_s^2(Q^2)\right)V(n,Q_0^2)\;.
\end{eqnarray}
Where $V$ is the valence quark compositions as
\begin{eqnarray}
\label{eq:VComposi} V(n,Q_0^2)=u_v(n,Q_0^2)+d_v(n,Q_0^2)\;.
\end{eqnarray}
As we see in Mellin-$n$ space the non-singlet (NS) parts of
structure function in the NNLO approximation for example, \ie
 $\;{\cal{M}}_n^{NNLO}(Q^2)$, can be obtained from
 the corresponding Wilson coefficients $C^{(k)}$  and the
  non-singlet quark
 densities.

In next section we will introduce the functional form of the
valence quark distributions and we will parameterize these
distributions at the scale of $Q_0^2$.

 By using the anomalous dimensions in one , two and three loops
from  \cite{Moch:2004pa} and inserting them in
Eq.~(\ref{eq:mnsNNLO}), the moment of non-singlet structure function
in the NNLO as a function of $n$ and $Q^2$ is available.

\section{Parametrization of the parton densities}
In this section we will discuss how we can determine the parton
distribution at the input scale of $Q_0^2=1$ GeV$^2$. To start the
parameterizations of the above mentioned parton distributions at the
input scale of $Q_0^2$ we assume the following functional form
\begin{equation}
\label{eq:xuv}
xu_{v}(x,Q_0^2)=N_{u}x^{a}(1-x)^{b}(1+c\sqrt{x}+dx)\;,
\end{equation}
\begin{equation}
\label{eq:xdv}
xd_{v}(x,Q_0^2)=\frac{N_{d}}{N_{u}}(1-x)^{e}\;xu_{v}(x,Q_0^2)\;.
\end{equation}

In the above the  $x^{a}$ term  controls the low-$x$ behavior parton
densities, and $(1-x)^{b,e}$ large values of $x$. The remaining
polynomial factor accounts for additional medium-$x$ values.
Normalization constants $N_u$ and $N_d$ are fixed by

\begin{equation}
\label{eq:const1} \int_0^1u_v(x)dx=2\;,
\end{equation}
\begin{equation}
\label{eq:const2} \int_0^1d_v(x)dx=1\;.
\end{equation}
The above normalizations are very effective to control unknown
parameters in Eqs.~(\ref{eq:xuv},\ref{eq:xdv}) via the fitting
procedure. The five parameters with $\Lambda _{QCD}^{N_{f}=4}$ will
be extracted by using the Bernstain polynomials
approach.\\

Using the valence quark distribution functions, the moments of
 $u_{v}(x,Q_0^2)$
and $d_{v}(x,Q_0^2)$ distributions can be easily calculated. Now
by inserting the Mellin moments of $u_v$ and $d_v$ valence quark
in the Eq.~(\ref{eq:VComposi}), the function of $V(n,Q_0^2)$
involves some unknown parameters.

\section{Averaged structure functions}
 Although it is relatively easy to compute the $n$th moment from the
 structure functions, the inverse process is not obvious. To do
 this inversion, we adopt a mathematically rigorous but easy
 method \cite{Yndu:78} to invert the moments and retrieve the structure
 functions. The method is based on the fact that for a given
 value of $Q^2$, only  a limited number of experimental points,
 covering a partial range of values of $x$ are available.  The method devised to deal with this situation
 is to take  averages of the structure function weighted by suitable
 polynomials. These  are the Bernstein polynomials which are defined by
 \be
B_{nk}(x)=\frac{\Gamma (n+2) }{\Gamma (k+1) \Gamma (n-k+1)
}x^k(1-x)^{n-k}\;;\; n\geq{k}\;. \ee Using the binomial expansion,
the above equation can be written as \be B_{n,\; k}(x)=\frac{\Gamma
(n+2)}{\Gamma (k+1)}\sum_{l=0}^{n-k}
\frac{(-1)^l}{l!(n-k-l)!}\;x^{k+l}\;. \label{eq:Bernnk} \ee
 Now, we can compare theoretical
predictions with the experimental results for the Bernstein
averages, which are given by \cite{Max:2002,KAM:2004} \be
F_{nk}(Q^2){\equiv}\int_{0}^{1}dxB_{nk}(x)F_3(x,Q^2)\;.
\label{eq:fnk1} \ee Therefore, the integral Eq.~(\ref{eq:fnk1})
represents an average of the function $F_3(x, Q^2)$ in the region
${\bar{x}}_{nk}-\frac{1}{2}\Delta{x}_{nk}{\leq}x{\leq}{\bar{x}}_{n,k}+
\frac{1}{2}\Delta{x}_{nk}$. The key point is, the values of $F_3$
outside this interval have a small contribution to the above
integral, as $B_{nk}(x)$ tends to zero very quickly. In order to
ensure the equivalence of the integral Eq.~(\ref{eq:fnk1}) to the
same integral in the range
$x_1={\bar{x}}_{nk}-\frac{1}{2}\Delta{x}_{nk}$ to
$x_2={\bar{x}}_{nk}+\frac{1}{2}\Delta{x}_{nk}$, we have to use the
normalization factor, $\int_{x_1}^{x_2}dxB_{nk}(x)$ in the
denominator of Eq.~(\ref{eq:fnk1}) which obviously is not equal to
$1$. So it can be written: \be
F_{nk}(Q^2){\equiv}\frac{\int_{{\bar{x}}_{nk}-
\frac{1}{2}\Delta{x}_{nk}}^{{\bar{x}}_{nk}+\frac{1}{2}\Delta{x}_{nk}}\;dx\;
B_{nk}(x)\;F_3(x,Q^2)} {\int_{{\bar{x}}_{nk}
-\frac{1}{2}\Delta{x}_{nk}}^{{\bar{x}}_{nk}+\frac{1}{2}\Delta{x}_{nk}}\;
dx\;B_{nk}(x)}\;. \ee By a suitable choice of $n$, $k$ we manage to
adjust to the region where the average is peaked around values which
we have
experimental data \cite{CCFR:1997}.\\

Substituting  Eq.~(\ref{eq:Bernnk}) in Eq.~(\ref{eq:fnk1}), it
follows that the averages of $F_3$ with $B_{nk}(x)$ as weight
functions can be obtained in terms of odd and even moments,
 \ba
F_{nk}&=&\frac{{(n-k)!}{\Gamma(n+2)}}{\Gamma(k+1)\Gamma(n-k+1)}\times
\nonumber
\\ &&\sum_{l=0}^{n-k}
\frac{(-1)^l}{l!(n-k-l)!}\;{{\cal{M}}({(k+l)+1}},Q^2)\;. \nonumber \\
\label{eq:FnkExpandMOM}\ea

We can only include a Bernstein average, $F_{nk}$, if we have
experimental points covering the whole range
 [${\bar{x}}_{nk}-\frac{1}{2}\Delta{x}_{nk}, {\bar{x}}_{nk}+\frac{1}{2}\Delta{x}_{nk}$]
 \cite{SantYun}. This means that with the available experimental data we can only use
 the following 28 averages, including\\

\begin{center}
$F_{2,1}^{(\exp )}(Q^{2}),F_{3,1}^{(\exp )}(Q^{2}), F_{4,2}^{(\exp
)}(Q^{2})$,..., $F_{13,4}^{(\exp )}(Q^{2})$.
\end{center}

Another restriction  we assume here, is  to ignore the effects of
moments with high order $n$ which do not strongly constrain the
fits. To obtain these experimental averages from CCFR data
\cite{CCFR:1997}, we fit $x{F_3}(x,{Q^2})$ for each bin in ${Q}^{2}$
separately to the convenient phenomenological expression \be
{xF_{3}}^{\hspace{-.12cm}{(phen)}}={\cal{A}}x^{\cal{B}}(1-x)^{\cal{C}}\;.
\label{eq:xf3pheno} \ee

\begin{figure}[tbh]
\centerline{\includegraphics[width=0.4\textwidth]{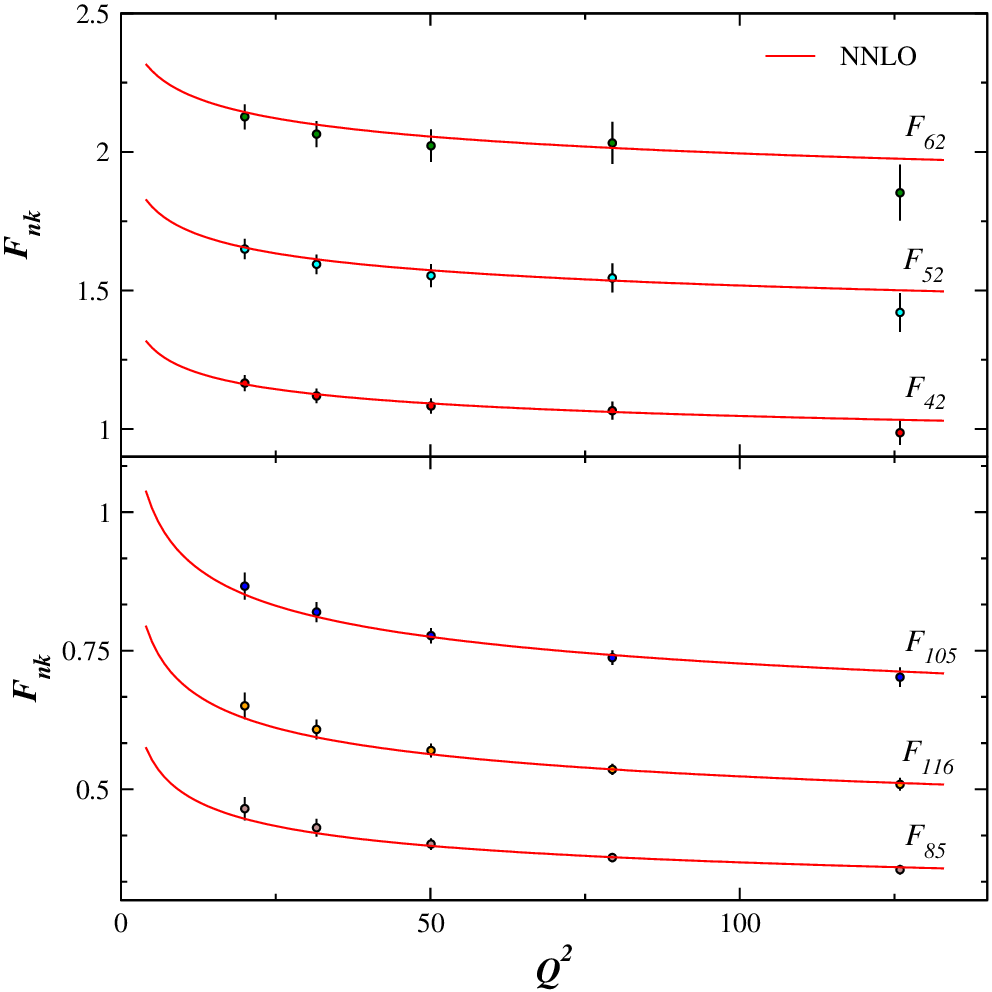}}
\caption{NNLO fit to Bernstein averages of $xF_3$.}
\label{pic:Fig2-FNKFIT}
\end{figure}

This form ensures zero values for ${xF_{3}}$ at $x=0$, and $x=1$. In
Table 1 we have presented  the numerical values of $\cal{A},\cal{B}$
and $\cal{C}$ at $Q^2=20, 31.6, 50.1, 79.4, 125.9$ GeV$^2$.
\begin{center}
\begin{tabular}{|c|ccc|}
\hline\hline
$Q^{2}(GeV^{2})$ & $\mathcal{A}$ & $\mathcal{B}$ & $\mathcal{C}$ \\
\hline\hline
$20$ & $4.742$ & $0.636$ & $3.376$ \\
$31.6$ & $5.473$ & $0.694$ & $3.659$ \\
$50.1$ & $5.679$ & $0.698$ & $3.839$ \\
$79.4$ & $4.508$ & $0.567$ & $3.757$ \\
$125.9$ & $7.077$ & $0.819$ & $4.246$ \\ \hline\hline
\end{tabular}
{\normalsize \\
 \textsf{\\Table~1: Numerical values of fitting
${\cal{A}},{\cal{B}},{\cal{C}}$ parameters in
Eq.~(\ref{eq:xf3pheno}). \label{tab1fitxf3}}}
\end{center}
Using Eq.~(\ref{eq:xf3pheno}) with the fitted values of
${\cal{A}},{\cal{B}}$ and ${\cal{C}}$, one can then compute
${F}_{nk}^{(exp)}({Q}^{2})$ in terms of Gamma functions. Some sample
experimental Bernstein averages are plotted in
Fig.~\ref{pic:Fig2-FNKFIT} in the higher approximations. The errors
in the ${F}_{nk}^{(exp)}(Q^2)$ correspond to allowing the CCFR data
for $x{F}_{3}$ to vary within the experimental error bars, including
the experimental systematic and statistical errors \cite{CCFR:1997}.
We have only included data for ${Q}^{2}{\ge}20{\rm{GeV}}^{2}$, this
has the merit of simplifying the analysis by avoiding evolution
through flavor thresholds.

Using Eq.~(\ref{eq:FnkExpandMOM}), the 28 Bernstein averages
${F}_{nk}({Q}^{2})$ can be written in terms of odd and even moments.
For instance: \ba
&&{F_{2,1}}(Q^2)=6\left({\cal{M}}(2,Q^2)-{\cal{M}}(3,Q^2)\right)\;,
\nonumber\\
&& \vdots \ea

The unknown parameters according to Eqs.~(\ref{eq:xuv},\ref{eq:xdv})
will be $a,b,c,d,e$ and $\Lambda _{QCD}^{N_{f}}$. Thus, there are 6
parameters for each order to be simultaneously fitted to the
experimental ${F}_{nk}({Q}^{2})$ averages. Using the CERN subroutine
MINUIT \cite{MINUIT:CERN}, we defined a global ${\chi}^{2}$ for all
the experimental data points  and found an acceptable fit with
minimum ${\chi}^{2}/{\rm{dof}}=74.772/134=0.558$ in the NNLO case
with the standard error of order $10^{-3}$. The best fit is
indicated by some sample curves in the Fig.~\ref{pic:Fig2-FNKFIT}.
The fitting
parameters and the minimum ${\chi}^{2}$ values in each order are listed in Table~2.\\

\begin{figure}[tbh]
\centerline{\includegraphics[width=0.4\textwidth]{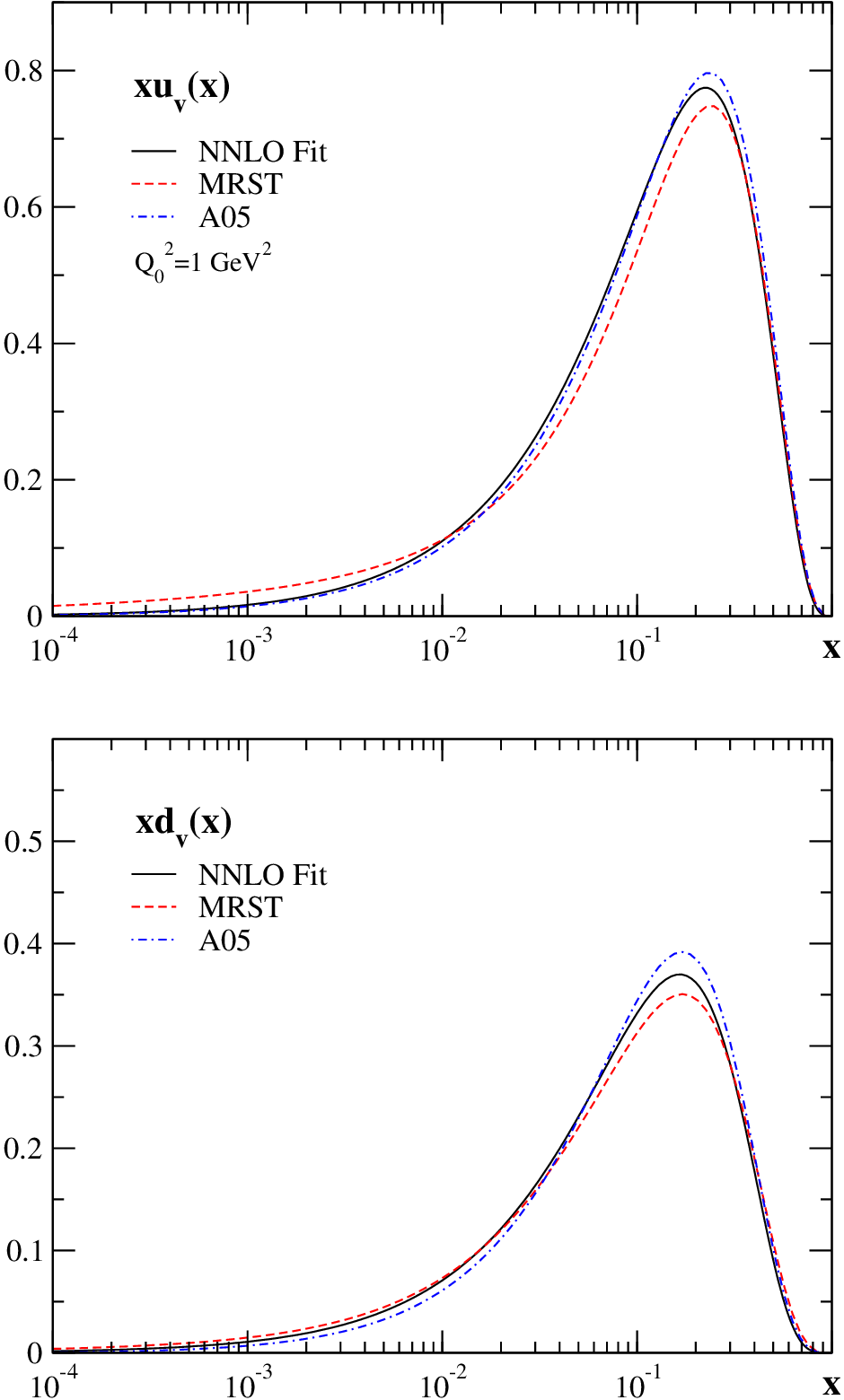}}
\caption{The parton densities $xu_v$ and $xd_v$ at the input scale
$Q_0^2=1.0\;GeV^2$ (solid line) compared to results obtained from
NNLO analysis by MRST (dashed line) \cite{Martin:2004} and
A05(dashed-dotted line)\cite{Alekhin:2005gq}.}
\label{pic:Fig3-partonQ0}
\end{figure}

From Eqs.(\ref{eq:xuv},\ref{eq:xdv}), we are able now to determine
the $xu_v$ and $xd_v$ at the scale of $Q_0^2$ in higher order
corrections. In Fig.~\ref{pic:Fig3-partonQ0} we have plotted the NLO
and NNLO approximation results of $xu_v$ and $xd_v$ at the input
scale $Q_0^2=1.0\;GeV^2$ (solid line) compared to the results
obtained from NNLO analysis (left panels) and NLO analysis (right
panels)  by MRST (dashed-dotted line) \cite{Martin:2004} and
A05(dashed
line)\cite{Alekhin:2005gq}.\\
\begin{center}
\begin{tabular}{|c|c|}
\hline\hline & NNLO \\ \hline\hline
$N_{u}$ & 5.134 \\
$a$ & 0.830 \\
$b$ & 3.724 \\
$c$ & 0.040 \\
$d$ & 1.449 \\ \hline\hline
$N_{d}$ & 3.348 \\
$e$ & 1.460 \\ \hline\hline $\Lambda _{QCD}^{(4)},MeV$ & 230 \\ \hline\hline $\chi ^{2}/$ndf & 0.558 \\
\hline\hline
\end{tabular}
 {\normalsize \\
 \textsf{\\Table~2: Parameter values of the NNLO non-singlet QCD fit at $Q_0^2=1$ GeV$^2$.}}
\end{center}

 All of the non-singlet parton distribution functions in
moment space for any order are now available, so we can use the
inverse Mellin technics to obtain the $Q^2$ evolution of valance
quark distributions which will be done in the next section.

\section{$\bf x$ dependent of valence quark densities}

In the previous section we parameterized the non-singlet parton
distribution functions at input scale of $Q_0^2=1$ GeV$^2$ in the
NNLO approximations by using Bernstein averages method. To obtain
the non-singlet parton distribution functions in $x$-space and for
$Q^2>Q_0^2$ GeV$^2$ we need to use the $Q^2$-evolution in $n$-space.
To obtain the $x$-dependence of parton
distributions from the $n-$%
dependent exact analytical solutions in the Mellin-moment space, one
has to perform a numerical integral in order to invert the
Mellin-transformation \cite{Graudenz:1995sk} \ba
 f^k(x,Q^{2})&=&\frac{1}{\pi }\int_{0}^{\infty }dw \times \nonumber \\
 &&Im[e^{i\phi}x^{-c-we^{i\phi }} M_k(n=c+we^{i\phi },Q^{2})]\;,\nonumber \\ \ea
where the contour of the integration lies on the right of all
singularities of $ M_k(n=c+we^{i\phi },Q^{2})$ in the complex
$n$-plane. For all practical purposes one may choose $c\simeq 1,\phi
=135^{\circ }$ and an upper limit of integration, for any $Q^{2}$,
of about $5+10/\ln x^{-1}$, instead of $\infty $, which guarantees
stable numerical results \cite{GRV:90,GRV:92}.
\section{Conclusion}
\label{sec:alphaMZ}

\begin{figure}[tbh]
\centerline{\includegraphics[width=.5\textwidth]{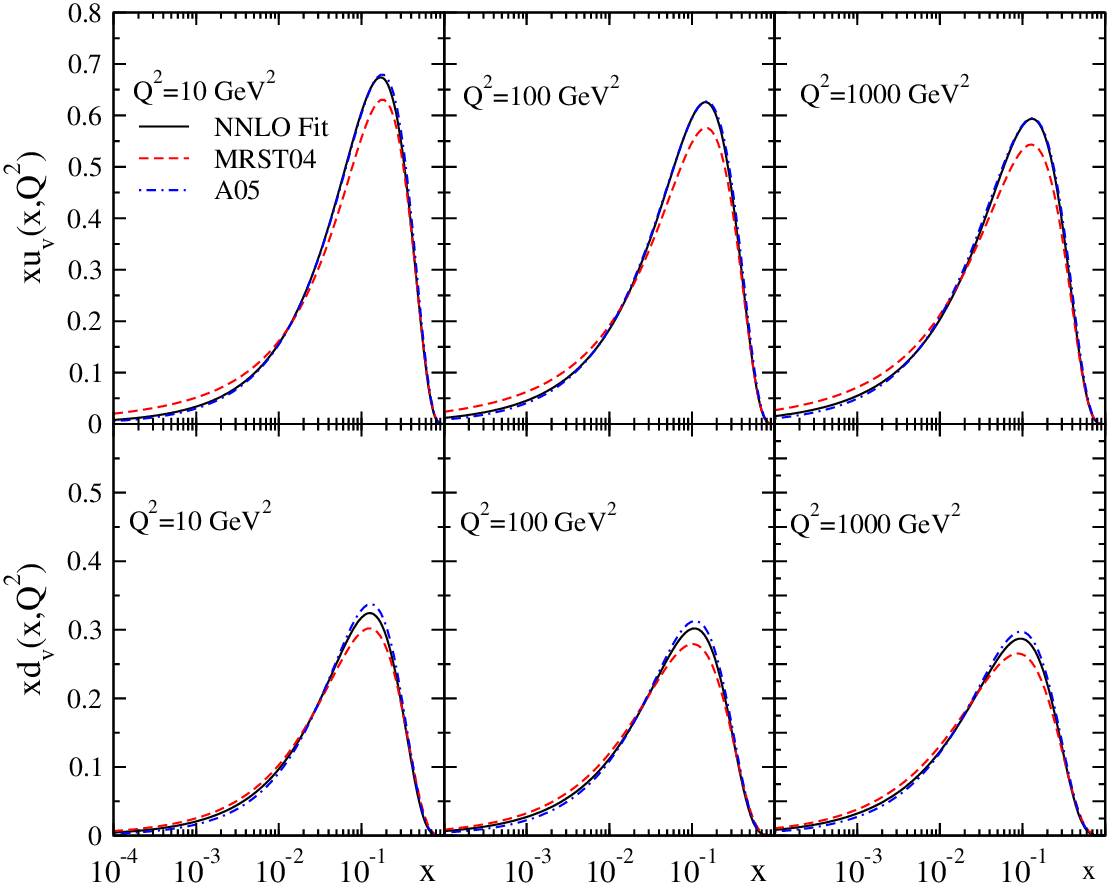}}
\caption{The parton distribution $xu_v$ and $xd_v$ at
  some different values of $Q^{2}$. The solid line is our model, dashed line is
  the MRST model
\cite{Martin:2004}, dashed-dotted line is the model from
\cite{Alekhin:2005gq}.} \label{pic:Fig4-xuvxdv}
\end{figure}

 The QCD analysis is performed in NNLO based on Bernstein polynomial approach. We determine the
valence quark densities in a wide range of $x$ and $Q^2$.
Inserting the functions of $q_{v}(n,Q^2)$ for  $q=u,d$ in Eq. (22)
we can obtain all valence distribution functions in fixed $Q^2$
and in $x$-space. In Fig.~\ref{pic:Fig4-xuvxdv} we have presented
the parton distribution $xu_v$ at
  some different values of $Q^{2}$. These distributions were compared  with some theoretical
models \cite{Martin:2004,Alekhin:2005gq}.\\
In Fig.~\ref{pic:Fig4-xuvxdv} we have also presented the same
distributions for $xd_v$.

The QCD scale $\Lambda_{\rm QCD}^{\rm N_f =4}$ is determined
together with the parameters of the parton distributions.  In our
fit results the value of $\Lambda_{\rm QCD}^{\rm N_f =4}$ and
$\alpha_s(M_Z^2)$ at the NNLO analysis is  $230$ MeV and $0.1142$
respectively.\\
Complete details of this paper with calculation of LO, NLO and NNLO
and also comparing them with together is reported in Ref. [36].

\section{Acknowledgments}
We are grateful to A. Mirjalili  for useful suggestions and
discussions. A.N.K acknowledge to Semnan university for the
financial support of this project.


\begin{thebibliography}{99}
\bibitem{Gross:1973rr}
D.J. Gross and F. Wilczek,
  Phys. Rev. D8 (1973) 3633.

\bibitem{Georgi:1974sr}
H. Georgi and H.D. Politzer,
  Phys. Rev. D9 (1974) 416.

\bibitem{Altarelli:1977zs}
G. Altarelli and G. Parisi,
  Nucl. Phys. B126 (1977) 298.

\bibitem{Floratos:1977au}
E.G. Floratos, D.A. Ross and C.T. Sachrajda,
  Nucl. Phys. B129 (1977) 66.

\bibitem{Floratos:1979ny}
E.G. Floratos, D.A. Ross and C.T. Sachrajda,
  Nucl. Phys. B152 (1979) 493.

\bibitem{Gonzalez-Arroyo:1979df}
A. Gonzalez-Arroyo, C. Lopez and F.J. Yndurain,
  Nucl. Phys. B153 (1979) 161.

\bibitem{Gonzalez-Arroyo:1980he}
A. Gonzalez-Arroyo and C. Lopez,
  Nucl. Phys. B166 (1980) 429.

\bibitem{Curci:1980uw}
G. Curci, W. Furmanski and R. Petronzio,
  Nucl. Phys. B175 (1980) 27.

\bibitem{Furmanski:1980cm}
W. Furmanski and R. Petronzio,
  Phys. Lett. 97B (1980) 437.

\bibitem{Floratos:1981hs}
E.G. Floratos, C. Kounnas and R. Lacaze,
  Nucl. Phys. B192 (1981) 417.

\bibitem{Hamberg:1992qt}
R. Hamberg and W.L. van Neerven,
  Nucl. Phys. B379 (1992) 143.

\bibitem{vanNeerven:1991nn}
W.L. van Neerven and E.B. Zijlstra,
  Phys. Lett. B272 (1991) 127.

\bibitem{Zijlstra:1991qc}
E.B. Zijlstra and W.L. van Neerven,
  Phys. Lett. B273 (1991) 476.

\bibitem{Zijlstra:1992kj}
E.B. Zijlstra and W.L. van Neerven,
  Phys. Lett. B297 (1992) 377.

\bibitem{Zijlstra:1992qd}
E.B. Zijlstra and W.L. van Neerven,
  Nucl. Phys. B383 (1992) 525.

\bibitem{Moch:2004pa}
 S.~Moch, J.~A.~M.~Vermaseren and A.~Vogt,
  Nucl.\ Phys.\ B {\bf 688} (2004) 101.

\bibitem{Vogt:2004ns}
  A.~Vogt, Comput.\ Phys.\ Commun.\  {\bf 170}, (2005) 65.


\bibitem{CCFR:1997}
  W.~G.~Seligman {\it et al.}, Phys.\ Rev.\ Lett.\  {\bf 79},  (1997) 1213.



\bibitem{Kataev:1997nc}
  A.~L.~Kataev, A.~V.~Kotikov, G.~Parente and A.~V.~Sidorov,
  Phys.\ Lett.\ B {\bf 417}, (1998) 374.

\bibitem{Kataev:1998ce}
  A.~L.~Kataev, G.~Parente and A.~V.~Sidorov,
  arXiv:hep-ph/9809500.

\bibitem{Alekhin:1998df}
  S.~I.~Alekhin and A.~L.~Kataev,
   Phys.\ Lett.\ B {\bf 452}, (1999) 402.




\bibitem{Kataev:1999bp}
  A.~L.~Kataev, G.~Parente and A.~V.~Sidorov,
    Nucl.\ Phys.\ B {\bf 573}, (2000) 405.




\bibitem{Kataev:2001kk}
  A.~L.~Kataev, G.~Parente and A.~V.~Sidorov,
     Phys.\ Part.\ Nucl.\  {\bf 34},  (2003) 20; A.~L.~Kataev, G.~Parente and A.~V.~Sidorov, Nucl.\ Phys.\ Proc.\
Suppl.\  {\bf 116} (2003) 105.


\bibitem{Santiago:2001mh}
  J.~Santiago and F.~J.~Yndurain,
  Nucl.\ Phys.\ B {\bf 611},  (2001) 447.





\bibitem{Moch:1999eb}
  S.~Moch and J.~A.~M.~Vermaseren,
  Nucl.\ Phys.\ B {\bf 573},  (2000) 853.




\bibitem{Yndu:78}
 F.~J.~Yndurain, Phys.\ Lett.\ B {\bf 74} (1978) 68.


\bibitem{Max:2002}
C.~J.~Maxwell and A.~Mirjalili,  Nucl.\ Phys.\ B {\bf 645} (2002)
298.


\bibitem{KAM:2004} Ali N. Khorramian, A. Mirjalili, S. Atashbar Tehrani,
{\it JHEP}{\bf 10}, 062 (2004).




\bibitem{SantYun}
 J.~Santiago and F.~J.~Yndurain,
    Nucl.\ Phys.\ B {\bf 563} (1999) 45
;

 J.~Santiago and F.~J.~Yndurain,
     Nucl.\ Phys.\ B {\bf 611} (2001) 447


\bibitem{MINUIT:CERN} F. James, CERN Program Library Long Writeup D506.



\bibitem{Martin:2004}
A.~D.~Martin, R.~G.~Roberts, W.~J.~Stirling and R.~S.~Thorne,
   Phys.\ Lett.\ B {\bf 604}, (2004) 61

\bibitem{Alekhin:2005gq}
  S.~Alekhin, JETP Lett.\  {\bf 82}, 628 (2005)
 ; S.~Alekhin,
 Phys.\
Rev.\ D {\bf 68} (2003) 014002.


\bibitem{Graudenz:1995sk}
  D.~Graudenz, M.~Hampel, A.~Vogt and C.~Berger,
    Z.\ Phys.\ C {\bf 70}, (1996) 77

\bibitem{GRV:90}
M.~Gluck, E.~Reya and A.~Vogt,
    Z.\ Phys.\ C {\bf 48} (1990) 471.

\bibitem{GRV:92}
  M.~Gluck, E.~Reya and A.~Vogt,
    Phys.\ Rev.\ D {\bf 45} (1992) 3986.

    \bibitem{KA:2006}
  Ali  N.~Khorramian, S. Atashbar Tehrani, hep-ph/0610136.




\end{thebibliography}
\end{document}